\documentclass[trackchanges]{aastex701}

\usepackage{amsmath}

\begin{document}

\title{Testing string theory with combined cosmological probes: a case study for dark matter gravitons
}

\author[0000-0001-9110-5550,sname=Pourtsidou,gname=Alkistis]{Alkistis Pourtsidou}
\affiliation{Institute for Astronomy, University of Edinburgh, \\ Royal Observatory, Blackford Hill, Edinburgh, EH9 3HJ, U.K.}
\affiliation{Higgs Centre for Theoretical Physics, School of Physics and Astronomy, \\ Edinburgh EH9 3FD, UK}
\email{alkistis.pourtsidou@ed.ac.uk}

\begin{abstract}
The string theory Swampland programme has resulted in distinct scenarios for the particle nature of dark matter. In this research note, I use combined cosmological probes to constrain the \emph{Dark Dimension} scenario, which predicts that dark matter consists of decaying massive gravitons characterised by a time-dependent kick velocity. I first provide updated constraints using a combination of CMB and BAO data. I then produce forecasts for a Stage-IV tomographic survey, and outline strategies for predicting the model's behaviour on nonlinear scales.
The results demonstrate the potential of surveys like Euclid and LSST to confirm or rule out string theory models of the dark sector.  
\end{abstract}


\section{Introduction and Formalism}

The \emph{Dark Dimension} model of the Universe \citep{Montero:2022prj} combines string theory principles (the ``Swampland'' programme) and the observational fact of the smallness of the cosmological constant, $\Lambda$, to predict one extra dimension in the micron range. The model leads to a natural dark matter candidate, the excitations of a tower of massive gravitons \citep{Gonzalo:2022jac}. 
The key feature is that these gravitons decay to lighter ones without losing significant total mass density, and impact a time-dependent ``kick velocity''
\begin{equation}
    v_{\rm kick} \propto t^{1/7} \, ,
\end{equation}
which affects structure formation, leading to suppressed structure growth. 

Using the fluid description in synchronous gauge, the continuity and Euler equations for the evolution of the dark graviton linear perturbations can be written as
\begin{equation}
\begin{aligned}
\dot{\delta} & =-(1+w)\left(\theta+\frac{\dot{h}}{2}\right)-3 \mathcal{H}\left(c_a^2-w\right) \delta \\
\dot{\theta} & =-\mathcal{H}\left(1-3 c_a^2\right) \theta+\frac{k^2 c_a^2}{1+w} \delta,
\end{aligned}
\end{equation}
where $w$ is the equation of state and $c_a^2$ the speed of sound. The expressions for these are (see \citet{Obied:2023clp} for details)
\begin{equation}
\begin{aligned}
w & \propto v_{\text {0}}^2\left(\frac{t}{t_{\text {today }}}\right)^{2 / 7} \\
c_a^2 & \simeq \frac{5w}{3(1+w)} \, ,
\end{aligned}
\end{equation}
where the $\Lambda$CDM limit is recovered for $v_0 \rightarrow 0$.

It is indeed $v_0$, the value of the kick velocity today, that can be constrained by cosmological observations. In the paper by \citet{Obied:2023clp}, a bound of $v_0 < 1.2 \times 10^{-3}c$ at 95\% C.L. was found using CMB data from Planck (PR3, \citet{Aghanim:2019ame}) and BAO data from eBOSS \citep{Cuceu:2019for}. This bound was tightened to $v_0 < 2.2 \times 10^{-4}c$ (which is $\sim 66$ km/s) adding cosmic shear data from KiDS-1000 \citep{Kuijken:2019gsa}, and assuming that purely linear cuts can be defined. 

\section{Methodology and Results}

In order to constrain the model with the latest CMB and BAO data, I implement it in the \texttt{CAMB} Boltzmann solver \citep{Lewis:1999bs}, and then I perform an MCMC analysis using \texttt{Cobaya} \citep{Torrado:2020dgo}. I vary $v_0$ and the standard cosmological parameters 
$\{\omega_{\text {b}}, \omega_{\text {cdm }}, \theta_{\rm s}, A_{\rm s}, n_{\rm s}, \tau_{\text {reio }}\}$, with fixed $\sum m_\nu = 0.06$ eV. The chains are analysed with \texttt{GetDist} \citep{Lewis:2019xzd}. I use the following datasets and likelihoods: \\
\textbf{Planck+ACT:} \emph{Planck} 2018 low-$\ell$ temperature and polarisation likelihood \citep{Aghanim:2019ame}, the \texttt{CamSpec} high-$\ell$ TTTEEE temperature and polarization likelihood using
\texttt{NPIPE} (\emph{Planck} PR4) \citep{Rosenberg:2022sdy}, and the combination of \emph{Planck} and ACT DR6 CMB lensing from  \cite{ACT:2023kun}. \\
\textbf{DESI:} Baryon Acoustic Oscillation (BAO) likelihood for all tracers from DESI DR2 \citep{DESI:2025zgx}. 
\begin{figure}[h]
    \centering    \includegraphics[width=0.5\columnwidth]{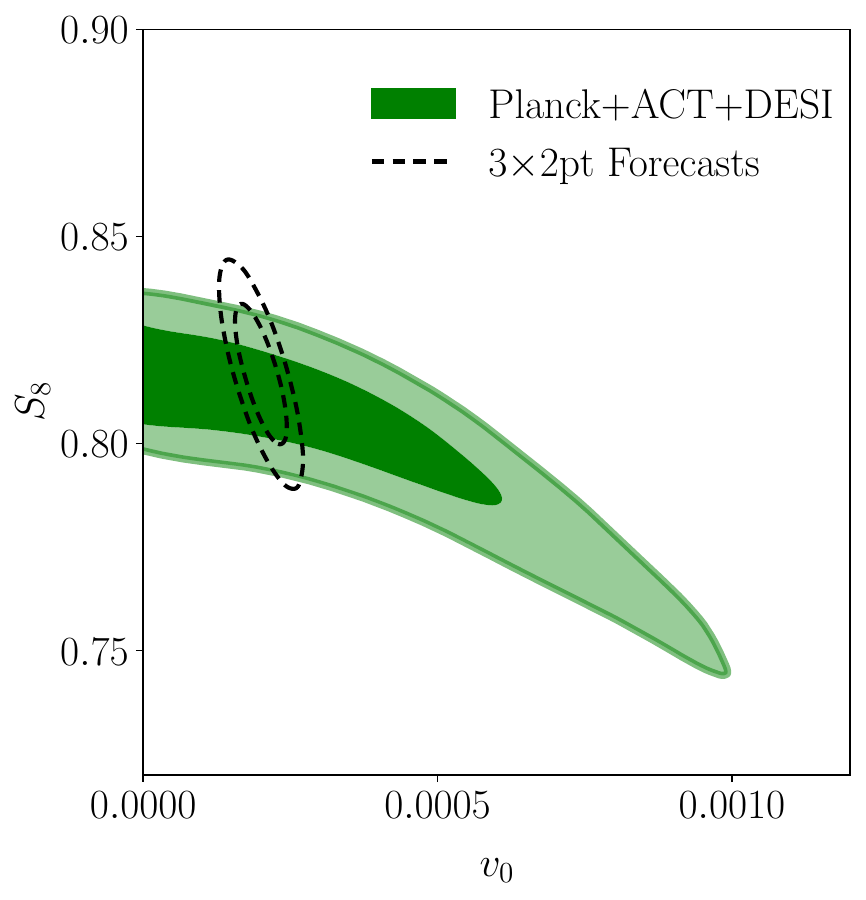}
    \caption{Contours containing 68\% and 95\% of the posterior probability for the ($S_8$,$v_0$) parameters. The plot shows constraints from the latest CMB and BAO data (solid green) as well as $3\times2$pt forecasts for an LSST Y1-like survey (dashed black).}
    \label{fig:GDMconstraints}
\end{figure}

The results of this analysis are shown in Figure~\ref{fig:GDMconstraints} (solid green). As expected \citep{Obied:2023clp}, there is a negative correlation between $S_8$ and $v_0$, as larger velocities correspond to more suppression. I find an upper bound $v_0 < 0.8 \times 10^{-3}c$ at 95\% C.L. -- this is tighter than the CMB$+$BAO bound in \citet{Obied:2023clp}, presumably due to the updated likelihoods and datasets used, and/or differences in the neutrino mass assumptions.
I then perform a Fisher Matrix forecast for a $3\times2$pt survey, assuming LSST Y1-like specifications \citep{LSSTDarkEnergyScience:2018jkl}. Such a survey requires accurate nonlinear modelling of the matter power spectrum, as well as modelling of baryonic feedback effects \citep{Robertson:2026ecl}. Defining appropriate scale cuts and dealing with the precision vs accuracy problem is a complicated challenge (see, for example, \citet{Truttero:2026bec}). I consider a common scale cut for all probes, $\ell_{\rm max}=1000$, and linear bias modelling. I apply a BBN prior on $w_b$, and I consider one baryonic parameter, $\mathrm{log}\,T_{\rm AGN}$ \citep{Mead:2020vgs}. For modelling the nonlinear matter power spectrum, I use the emulator from \citet{Bucko:2024izb}, which is based on a suite of $N$-body simulations. This emulator models the ratio of the nonlinear decaying dark matter power spectra to $\Lambda$CDM with 3 parameters: the decay rate, $\Gamma$, the velocity kick, $v_{\rm kick}$, and the fraction of decaying to total dark matter, $f$. In the model considered here, one can fix $f=1$ and $\Gamma \simeq 1/140$ Gyr$^{-1}$ \citep{Obied:2023clp}. The results are shown in Figure~\ref{fig:GDMconstraints} (dashed black), for a fiducial cosmology with $v_0=0.2 \times 10^{-3}c$. I find that $v_0$ can be constrained with a 15\% fractional error.  

These forecasts demonstrate the potential of surveys like LSST and Euclid \citep{Euclid:2024yrr} to confirm or rule out string theory based models of the dark sector, but accurate nonlinear modelling is crucial. Using this work as a case study, the nonlinear correction used is cosmology independent, which might not be an adequate assumption for LSST and Euclid's requirements. The way forward is to consider approaches such as the \emph{halo model reaction} \citep{Cataneo:2018cic, Bose:2020wch}, validate them with bespoke simulations, perform scale cut challenges such as the ones in \citet{Robertson:2026ecl,Truttero:2026bec}, reanalyse data from KiDS-1000 and DES \citep{DES:2026fyc}, and prepare the analysis of Euclid and LSST data. This is the subject of future work.

\begin{acknowledgments}
I thank Maria Tsedrik and Ben Bose for their help and feedback. My research is supported by a UK Research and Innovation Future Leaders Fellowship [grant MR/X005399/1].
\end{acknowledgments}

\bibliography{main}{}
\bibliographystyle{aasjournalv7}

\end{document}